\documentclass[twocolumn]{aastex61}
\submitjournal{ApJ}

\shorttitle{Suspended and restored activities of Super Massive Black Hole}
\shortauthors{Wang et al.}

\begin{document}


\title{Suspended and restored activities of A Nearby Super Massive Black Hole}

\correspondingauthor{J. Wang}
\email{wj@bao.ac.cn}
\correspondingauthor{D. W. Xu}
\email{dwxu@bao.ac.cn}

\author{J. Wang}
\affil{Guangxi Key Laboratory for Relativistic Astrophysics, School of Physical Science and Technology, Guangxi University,
Nanning 530004, People始s Republic of China}
\affil{Key Laboratory of Space Astronomy and Technology, National Astronomical Observatories, Chinese Academy of Sciences, Beijing
100101, China}

\author{D. W. Xu}
\affil{Key Laboratory of Space Astronomy and Technology, National Astronomical Observatories, Chinese Academy of Sciences, Beijing
100101, China}
\affil{School of Astronomy and Space Science, University of Chinese Academy of Sciences, Beijing, China}

\author{S. S. Sun}
\affil{Guangxi Key Laboratory for Relativistic Astrophysics, School of Physical Science and Technology, Guangxi University,
Nanning 530004, People始s Republic of China}
\affil{Key Laboratory of Space Astronomy and Technology, National Astronomical Observatories, Chinese Academy of Sciences, Beijing
100101, China}

\author{Q. C. Feng}
\affil{Key Laboratory of Space Astronomy and Technology, National Astronomical Observatories, Chinese Academy of Sciences, Beijing
100101, China}

\author{T. R. Li}
\affil{Key Laboratory of Optical Astronomy, National Astronomical Observatories, Chinese Academy of Sciences, Beijing
100101, China}

\author{P. F. Xiao}
\affil{Key Laboratory of Space Astronomy and Technology, National Astronomical Observatories, Chinese Academy of Sciences, Beijing
100101, China}

\author{J. Y. Wei}
\affiliation{Key Laboratory of Space Astronomy and Technology, National Astronomical Observatories, Chinese Academy of Sciences, Beijing
100101, China}
\affiliation{School of Astronomy and Space Science, University of Chinese Academy of Sciences, Beijing, China}



\begin{abstract}

The discovery of spectral type transition of active galactic nuclei (AGNs), the so-called “changing-look” (CL) phenomenon, challenges
the widely accepted AGN paradigm, not only in the orientation based Unified Model, but also in the standard disk model.
In past decades, only a couple of nearby repeat ``changing-look'' active galactic nuclei (CL-AGNs) have been identified.
Here we report spectroscopic observations of UGC 3223 over the course of 18 years, from 2001 onwards. 
Combining the spectrum taken in 1987 by Stirpe, we have witnessed its type transitions from $1.5\rightarrow2.0\rightarrow1.8$ over 32 years, 
and captured a long-lived (at least 10 years) thorough ``turn-off'' state with a spectrum typical of a Seyfert 2 galaxy. 
The long-term  thorough ``turn-off'' state probably suggests a once-dormant and an awakening central engine in UGC3223. 
We argue the (dis)appearance of the broad Balmer emission lines can be explained by the disk-wind BLR model given the evolution of the 
calculated Eddington ratio of accretion of the supermassive black hole.

\end{abstract}




\section{Introduction} \label{sec:intro}

Multi-epoch spectroscopy recently enables us to reveal a batch of the so-called ``changing-look'' active galactic nuclei (CL-AGNs) that show 
a spectral type transition between type I, intermediate type and type II within a time scale of an order of years or decades. 
So far, the CL phenomenon has been identified by spectroscopy in only $\sim$80 AGNs, including both 
``turn-on'' and ``turn-off'' transitions
(e.g., Shapovalova et al. 2010; Shappee et al. 2014; LaMassa et al. 2015; McElroy et al. 2016; 
Runnoe et al. 2016; Gezari et al. 2017; Yang et al. 2018; Ruan et al. 2016; MacLeod et al. 2010, 2016; Wang et al. 2018, 2019; 
Stern et al. 2018; Frederick et al. 2019; Yan et al. 2019; Trakhtenbrot et al. 2019). 
Based on the  Catalina Real-time Transient Survey, Graham et al. (2019) recently identified 111 ``changing-state'' quasars according to their 
optical and mid-infrared photometric behavior and spectroscopic change.

The observed CL phenomenon challenges the traditional understanding of AGNs in two points. At first,  
there is competing evidence in both middle infrared and polarization supporting that the optical CL phenomenon is due to a variation in accretion  
rate of a supermassive blackhole (SMBH) 
(e.g., Sheng et al. 2017, Yang et al. 2018; Wang et al. 2018, 2019b;  
Gezari et al. 2017; Macleod et al. 2019; Rumbaugh et al. 2018; Stern et al. 2018; Hutsemekers et al. 2019), 
rather than the orientation effect (e.g., Antonucci 1993).
Secondly, there is a viscosity crisis, in which the expected viscous timescale of optical emission coming from the outer accretion is 
larger than the timescale of the observed CL phenomenon by an order of magnitude (e.g., Lawrence 2018 and references therein).  
This crisis could be theoretically resolved by introducing a local disk thermal instability
(e.g., Husemann et al. 2016; Jiang et al. 2016)
or a magnetic field (e.g., Ross et al. 2018; Stern et al. 2018; Dexter \& Begelamn 2019).

Even though most of the CL-AGNs are identified in the case of two-epochs spectroscopy,
repeat type transitions have been identified in only a few nearby Seyfert galaxies and quasars: Mrk\,590, Mrk\,1018, NGC\,1566,
NGC\,4151, NGC\,7603, Fairall\,9 and 3C\,390.3 (Marin et al. 2019). Among these objects, only Mrk\,590 shows a thorough disappearance of its BLR emission 
(Denney et al. 2014).  
We here report a discovery of a new nearby CL-AGN, UGC\,3223, with repeat type transition and a prolonged thorough ``turn-off'' of the activity of the SMBH. 

%

The paper is organized as follows. Section 2 describes our repeat spectroscopic observations and data reductions. The spectral 
analysis is presented in Section 3. Section 4 presents the results and discussion. A $\Lambda$CDM cosmology with parameters
$H_0=70\mathrm{km\ s^{-1}\ Mpc^{-1}}$, $\Omega_m=0.3$, and
$\Omega_\Lambda=0.7$ is adopted throughout the paper.


\section{Observations and Data Reduction} \label{sec:style}

UGC\,3223 (=MCG +01-13-012, R.A.=04$^\mathrm{h}$59$^\mathrm{m}$09$^\mathrm{s}$.4, DEC=+04\degr58\arcmin30\arcsec, J2000) is a local AGN 
at redshift of $z=0.015621$. It was classified as a Seyfert 1.5 galaxy in the catalog of quasars and active nuclei: 12th edition
(Veron-Cetty \& Veron 2006), according to the spectroscopy taken in 1987 (Stirpe 1990).  
 
\subsection{Spectroscopic observations}

The long-slit spectra of UGC\,3223 have been taken by us by using the 2.16m telescope (Fan et al. 2016) at Xinglong  
observatory of National Astronomical Observatories, Chinese Academy of Sciences (NAOC) in four epochs from 2001 to 2019 (See Column(1) in Table 1).     
The observations were carried out with the Optomechanics Research Inc. spectrograph
equipped with a back-illuminated SPEC 1340$\times$400 CCD as detector. 
A grating of 300 $\mathrm{grooves\ mm^{-1}}$, and a slit of $2\arcsec$ oriented in the south-north direction 
were used in the observations. This setup finally leads to a spectral resolution of $\sim9$\AA\ as measured from the sky emission 
lines and comparison arcs. The blazed wavelength was fixed at 6000\AA\ in all the observation runs, which provides a wavelength coverage 
from 3400-8600\AA\ in the observer frame.  In each of the runs, the object was observed twice in
succession. The exposure time of each frame ranges from 1500 to 3600 seconds.
The wavelength and flux calibrations were carried out by the helium-neon-argon comparison arc taken between the
two successive frames, i.e., at the position nearly identical to that of the object, and by the Kitt
Peak National Observatory (KPNO) standard stars (Massey et al. 1988), respectively. 


\subsection{Data reduction}

We reduced the two-dimensional spectra by the standard procedures through the IRAF package\footnote{IRAF is distributed by the 
National Optical Astronomical Observatories,
which are operated by the Association of Universities for Research in
Astronomy, Inc., under cooperative agreement with the National Science
Foundation.}. At first, the bias subtraction and flat-field correction were applied to 
each observed frame.   
In each observation run, the two frames were combined prior to the extraction to enhance
the signal-to-noise ratio and to eliminate the contamination of cosmic rays easily.
Each extracted 1d spectrum was then calibrated in wavelength and flux by the corresponding comparison arc and standard. 
The two telluric features A and B around $\lambda$7600 and $\lambda$6800 due to the $\mathrm{O_2}$ molecules were 
removed from each extracted spectrum by the corresponding standard. 
The Galactic extinction was then corrected for each extracted spectrum by the extinction magnitude of $A_V=0.229$ (Schlafly \& Finkbeiner 2011) 
taken from the NASA/IAPC Extragalactic
Database (NED), assuming the $R_V = 3.1$ extinction law of our Galaxy (Cardelli et al. 1989). The spectra were then transformed
to the rest frame according to its redshift. 
The rest-frame spectra taken at the four different epochs are displayed in Figure 1 (the middle curve in each panel).

\begin{figure}
\plotone{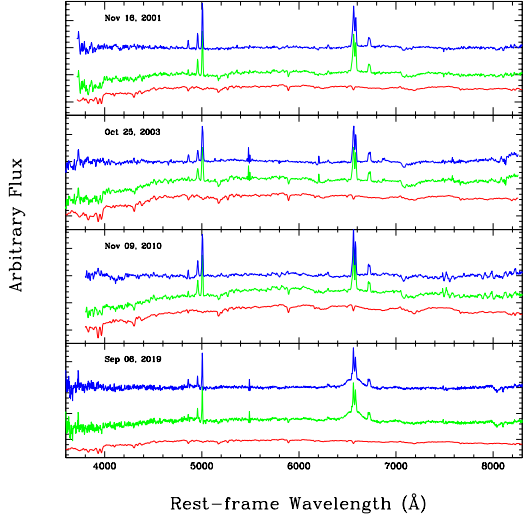}
\caption{The Xinglong spectra of UGC\,3223 taken in the four different epochs. In each panel, 
the starlight-subtracted spectrum,  the observed rest-frame spectrum and the modeled starlight component are 
displayed by the curves from top to bottom. 
}
\end{figure}

\section{Spectral Analysis} \label{subsec:tables}

One can see clearly from the figure a lack of broad H$\alpha$ emission line in the three Xinglong spectra taken in 2001, 2003 and 2011, which is different from 
the spectrum shown in Stirpe (1990). It is interesting that the broad H$\alpha$ emission and a weak 
H$\beta$ broad emission return back in the 2019 spectrum. A spectral analysis is performed 
as follows to quantify the type transition in UGC\,3223.

\subsection{Removing of The Star Light Component}

The starlight component of each of the four spectra is at first modeled by a linear combination of 
the first seven eigenspectra through a $\chi^2$ minimization. The eigenspectra are built from the
standard single stellar population spectral library developed by Bruzual \& Charlot (2003).  
An intrinsic extinction due to the host galaxy described by the Galactic extinction
curve with $R_V=3.1$ is also involved in our fitting. The minimization is carried out over
the rest-frame wavelength range from 3700 to 8000\AA,  except for the regions with strong emission lines,
e.g., Balmer lines (both narrow and potential broad components), [\ion{O}{3}]$\lambda\lambda$4959, 5007, [\ion{N}{2}]$\lambda\lambda$6548, 6583,
[\ion{S}{2}]$\lambda\lambda$6716, 6731, [\ion{O}{2}]$\lambda$3727, [\ion{O}{3}]$\lambda$4363, and [\ion{O}{1}]$\lambda$6300.
Because the absorption features in the observed spectra are dominated by the instrumental profile, 
the starlight templates are convolved with a fixed Gaussian profile before the minimization. 
The starlight component, along with the starlight-subtracted spectrum, are shown in Figure 1. 


\subsection{Line Profile Modeling}

In each starlight-subtracted spectrum, we model the emission line profiles by a linear combination of a set of Gaussian profiles in 
both H$\alpha$ and H$\beta$ regions by the SPECFIT task (Kriss 1994) in the IRAF package. 
The line flux ratios of the [\ion{O}{3}]$\lambda\lambda$4959, 5007 and the [\ion{N}{2}]$\lambda\lambda$6548, 6583
doublets are fixed to their theoretical values, i.e., 1:3. 
The line modelings are shown in the left and right panels of Figure 2  for the H$\beta$ and H$\alpha$ regions, respectively. 
In all the spectra except for the one taken in 2019,  each Balmer line 
profile can be well reproduced by a corresponding narrow component.  
However, a broad component is necessary for fitting the 
H$\alpha$ line profile of the 2019 spectrum. In addition, a weak H$\beta$ broad component is also required to properly reproduce the 2019
observation. 
The results of the profile modeling, along with the results quoted from Stirpe (1990) for the 1987 spectrum, are listed in Table 1. 
All the errors reported in the table correspond to the 1$\sigma$ significance level, where only the uncertainties caused by the fitting are included. 

\begin{figure}
\plotone{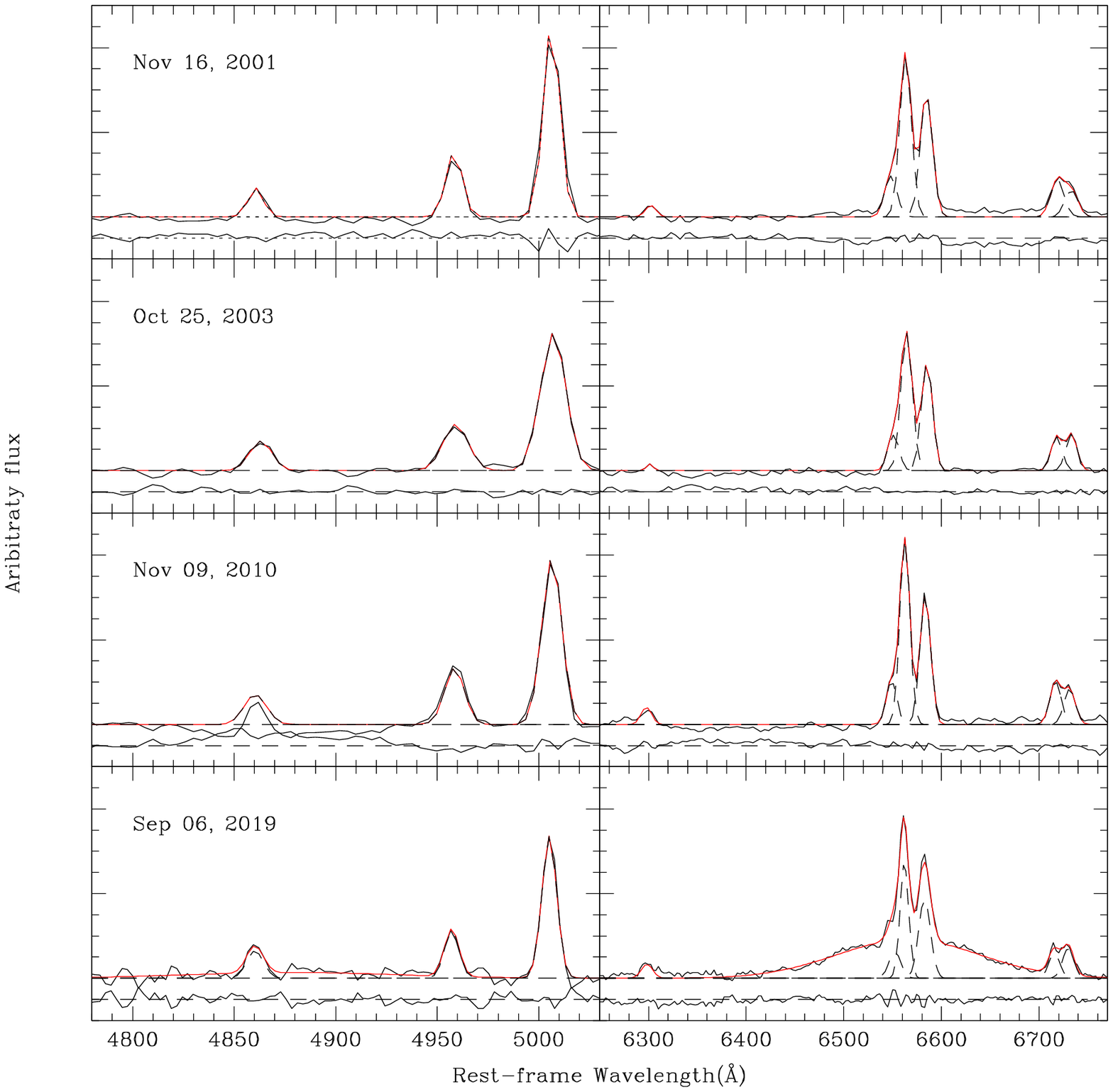}
\caption{An illustration of line profile modelings by a linear combination of a set of Gaussian functions for the H$\beta$ (the left panels) and 
H$\alpha$ (the right panels) regions. In each panel, the modeled continuum has already been removed from the original observed spectrum. 
The observed and modeled line profiles are plotted by black and red solid lines, respectively. Each Gaussian function is shown by a
dashed line. The sub-panel underneath the line spectrum presents the residuals between the observed and modeled profiles. 
}
\end{figure}

\begin{table*}[h!]
\renewcommand{\thetable}{\arabic{table}}
\centering
\caption{Spectral measurements and analysis.} 
\label{tab:decimal}
\tiny
\begin{tabular}{lccccccccc}
\tablewidth{0pt}
\hline
\hline
Date &  $F(\mathrm{H\beta_n})$ & $F(\mathrm{[OIII]\lambda5007})$ &  $F(\mathrm{H\alpha_n})$ &  $F(\mathrm{H\beta_b})$ & FWHM(H$\beta_\mathrm{b}$) & $F(\mathrm{H\alpha_b})$ & FWHM(H$\alpha_\mathrm{b}$) & $M_{\mathrm{BH}}/M_\odot$ & $L/L_{\mathrm{Edd}}$\\ 
     &  \multicolumn{4}{c}{$\mathrm{10^{-15}\ erg\ s^{-1}\ cm^{-2}}$} &   $\mathrm{km\ s^{-1}}$ & $\mathrm{10^{-15}\ erg\ s^{-1}\ cm^{-2}}$ & $\mathrm{km\ s^{-1}}$
     &    &  \\
(1)  &   (2) & (3) & (4) & (5) & (6) & (7) & (8) & (9) & (10) \\
\hline
1987/01/14$^a$  & 12.0         & 47.9         & 33.0         &  85.1     & 4740     & 275.4    & 3980  & $2.1\times10^7$ & 0.014\\
2001/11/16  & $6.5\pm1.6$  & $46.3\pm3.5$ & $56.5\pm6.4$ &  \dotfill & \dotfill & \dotfill & \dotfill  & \dotfill & \dotfill \\
2003/10/25  & $9.6\pm1.2$  & $51.1\pm2.4$ & $47.1\pm4.4$ &  \dotfill & \dotfill & \dotfill & \dotfill  & \dotfill & \dotfill \\
2010/11/09  & $9.8\pm2.0$  & $48.3\pm5.5$ & $56.1\pm4.8$ &  \dotfill & \dotfill & \dotfill & \dotfill  & \dotfill & \dotfill \\
2019/09/06  & $6.7\pm1.6$  & $30.9\pm2.5$ & $30.5\pm3.2$ &  $17.6\pm1.6$ & $7600\pm2600$ & $167.5\pm5.0$ &  $7800\pm200$ & $(9.8\pm2.8)\times10^7$ & $0.004\pm0.001$ \\
\hline
\hline
\end{tabular}
\tablenotetext{a}{The data are quoted from Stirpe (1990).}
\end{table*}

\subsection{Blackhole mass and Eddington ratio}

We then estimate the SMBH viral mass ($M_{\mathrm{BH}}$) and Eddington ratio $L_{\mathrm{bol}}/L_{\mathrm{Edd}}$  (where $L_{\mathrm{Edd}}=1.26\times10^{38}M_{\mathrm{BH}}/M_\odot\ \mathrm{erg\ s^{-1}}$ 
is the Eddington luminosity) of UGC\,3223 from the modeled broad H$\alpha$ emission lines, according to 
several well-established calibrated relationships (e.g., Kaspi et al. 2000, 2005; Marziani \& Sulentic 2012; 
Peterson 2014; Du et al. 2014, 2015).       

The calibration given by Green \& Ho (2007)
\begin{equation}
  M_{\mathrm{BH}}=3.0\times10^6\bigg(\frac{L_{\mathrm{H\alpha}}}{10^{42}\ \mathrm{erg\ s^{-1}}}\bigg)^{0.45}
  \bigg(\frac{\mathrm{FWHM(H\alpha)}}{1000\ \mathrm{km\ s^{-1}}}\bigg)^{2.06}M_\odot
\end{equation}
is used to estimate $M_{\mathrm{BH}}$. To obtain an estimation of $L_{\mathrm{bol}}/L_{\mathrm{Edd}}$, we derive the bolometric luminosity
$L_{\mathrm{bol}}$ from the standard bolometric correction $L_{\mathrm{bol}}=9\lambda L_\lambda(5100\AA)$ (e.g., Kaspi et al. 2000),
where $L_\lambda(5100\AA)$ is the AGN's specific continuum luminosity at 5100\AA. This specific luminosity can be inferred from
the H$\alpha$ broad-line luminosity through the calibration (Greene \& Ho 2005)
\begin{equation}
 \lambda L_\lambda(5100\AA)=2.4\times10^{43}\bigg(\frac{L_{\mathrm{H\alpha}}}{10^{42}\ \mathrm{erg\ s^{-1}}}\bigg)^{0.86}
\end{equation}

The estimated $M_{\mathrm{BH}}$ and $L_{\mathrm{bol}}/L_{\mathrm{Edd}}$ are tabulated in Columns (9) and (10) in Table 1, along with the uncertainties, respectively, after taking into account both
the proper error propagation and the uncertainties of the used calibrations\footnote{The uncertainties can not be determined for the 1987 spectrum because  of the lack of errors of the measured parameters in Stirpe (1990).}. 
In the estimation of $L_{\mathrm{H\alpha}}$, the intrinsic extinction has been corrected 
from the narrow-line ratio H$\alpha$/H$\beta$, assuming the Balmer decrement of standard case B recombination and a Galactic extinction curve with $R_V = 3.1$.

An extremely low $L_{\mathrm{bol}}/L_{\mathrm{Edd}}=0.004\pm0.001$ is finally found from the 2019 spectrum. 

\section{Results and Discussion}

The local AGN UGC\,3223, which was originally classified as a Seyfert 1.5 galaxy, has been spectroscopically observed by us in four different epochs in past 19 years. 
Our observations and spectral analysis indicate that the object changes to a Seyfert 2 galaxy without broad Balmer emission lines in the spectra taken in
2001, 2003 and 2011. A reappearance of the broad H$\alpha$ emission line (and also of a weak H$\beta$ broad emission) is found in our 2019 spectrum, 
although the line width of the broad H$\alpha$ in 2019 is about twice of the value obtained from the 1987 spectrum (Stirpe 1990).

We argue that the lack of broad Balmer component in the 2001, 2003 and 2010 spectra is not due to their low spectral resolution. 
First,  
the Balmer lines in the three spectra have measured line widths close to those of the forbidden narrow emission lines.
Second, the average narrow-line ratio of $F(\mathrm{H\alpha+NII\lambda\lambda6548,6583})/F(\mathrm{SII\lambda\lambda6716,6731})$ is 4.21 for the three spectra, 
which is highly consistent with the value of 4.19 of the 2019 spectrum. This consistency indicates that there is unlikely a potential broad component 
in the modeled line profiles of the three spectra.

Figure 3 shows the 
multi-wavelength light curves of UGC\,3223. Within the epochs of the last two spectra, its mid-infrared (MIR) brightness detected by the Wide-field 
Infrared Survey Explorer (WISE and NEOWISE-R, Wright et al. 2010; Mainzer et al. 2014) increases gradually when the object changes from a 
type-II AGN to a type-1.8 AGN, which supports the scenario that the type transition is likely due to the enhancement of accretion rate  
rather than the obscuring effect (e.g., (e.g., Sheng et al. 2017; Stern et al. 2018; Wang et al. 2019b).

Even though a breath of broad-line region (BLR) has been previously discovered in a few local AGNs by long term monitoring,
a thorough disappearance of the BLR emission is still rare.    
To our best knowledge, Mrk\,590 was the first, and maybe the only repeat CL-AGN that shows ``on-off-on'' transition sequence  
with a thorough ``turn-off''. The spectrum taken in 2014 indicates that Mrk\,590 finally changed to a true Seyfert 2 galaxy from a 
typical type-I AGN (Denney et al. 2014). After a period of a few years, the object awakes and turns back a type-I AGN 
(Mathur et al. 2018; Raimundo et al. 2019).  
Similar to Mrk\,590 and UGC\,3223, Mrk\,1018 changes from type 1.9 to type 1.0 in the spectrum taken in 1980s (Cohen et al. 1986), and changes back 
type 1.9 in the 2016 spectrum (Husemann et al. 2016; McElory et al. 2016).

%

Figure 4 compares UGC\,3223 to whole AGN population and to a compilation of known CL-AGNs (Wang et al. 2019b) by marking it
on the distributions of $L_{\mathrm{bol}}-M_{\mathrm{BH}}$ (the left panel) and $L_{\mathrm{bol}}-L/L_{\mathrm{Edd}}$ (the right panel). 
UGC\,3223 marginally stands out of the known CL-AGNs due to the lowest $L/L_{\mathrm{Edd}}\sim0.004$.
SDSS\,J233602.98+001728.7 at $z=0.243$ is located adjacent to UGC\,3223 in both distributions. It is interesting that
SDSS\,J233602.98+001728.7 shows similarity to UGC3223 in type transition from type-1.5 to type-2 (Ruan et al. 2016), 
although its repeat transition has not yet been discovered.

Two scenarios can potentially account for the discovered repeat type transition with a thorough ``turn-off'' exhibition.
On the one hand, the disappearance and re-appearance of the BLR can be understood in the context of the previously 
proposed disk-wind BLR models, because the $L_{\mathrm{bol}}/L_{\mathrm{Edd}}$ of UGC\,3223 is most close to the 
critical value predicted by the disk-wind models. A critical value of $L_{\mathrm{bol}}/L_{\mathrm{Edd}}\approx2-3\times10^{-3}$ 
is in fact proposed by Nicastro (2000) for a BH mass within a range of $10^{7-8}M_\odot$. 
This critical value is resulted from a critical radius of the accretion disk where the power deposited into the vertical outflow is maximum.
In the model proposed in Elitzur \& Ho (2009), an observable BLR can not be sustained below a certain 
luminosity $L\approx5\times10^{39}(M_{\mathrm{BH}}/10^7M_\odot)^{2/3}\ \mathrm{erg s^{-1}}$ because the mass outflow rate scales 
with $L$ as $L^{1/4}$ (Elitzur \& Shlosman 2006). A much lower critical value of $L_{\mathrm{bol}}/L_{\mathrm{Edd}}\sim10^{-6}$ is therefore predicted for a transition between 
disappearance and appearance of BLR when the fiducal values of a set of parameters of the disk are adopted. In the
case of UGC\,3223, the critical values of disappearance of BLR
are $L\sim2\times10^{40}\ \mathrm{erg\ s^{-1}}$ and $L_{\mathrm{bol}}/L_{\mathrm{Edd}}\sim2\times10^{-6}$.   
On the other hand, 
an once-dormant and reactive central engine in UGC\,3223 can not be excluded at current stage, although the
physical reason of shutting down of fueling of the SMBH is still an open question.

\begin{figure}
\plotone{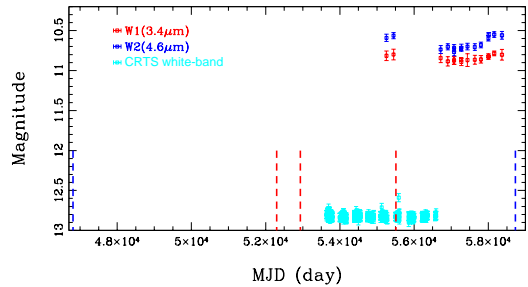}
\caption{The multi-wavelength light curves of UGC\,3223. For the Wise $w1$ and $w2$ bands, the 
light curves are binned by averaging the measurements within one day. The vertical dashed lines mark the epochs of spectra, where the blue lines
denote a spectrum with broad Balmer lines, and the red ones a spectrum without broad Balmer lines.
}
\end{figure}

\begin{figure}
\plotone{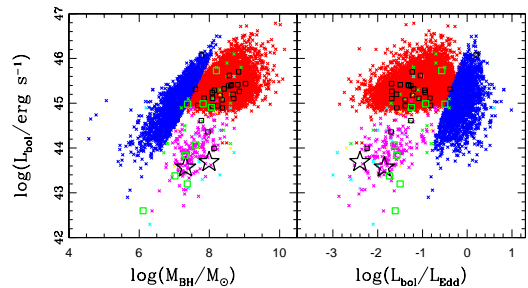}
\caption{The $L_{\mathrm{bol}}-M_{\mathrm{BH}}$ (the left panel) and $L_{\mathrm{bol}}-L/L_{\mathrm{Edd}}$(the right panel) diagrams.
UGC3223 is marked by the black open stars. The ``turn-on'' states of
known CL-AGNs and repeat CL-AGNs are denoted by the black and
green open squares, respectively. The used comparison samples are described
as follows. Red cross: the quasars with $z<0.5$ taken from the value-added SDSS
DR7 quasar catalog; blue cross: the SDSS DR3 NLS1 catalog; magenta
cross: the SDSS DR7 intermediate-type AGNs. The Swift/BAT AGN sample
is shown by the green, yellow and cyan cross for Seyfert 1, 1.2 and 1.5 galaxies, respectively. See Wang et al. (2019b) and references therein
for the details of the used comparison samples.
}
\end{figure}

\section{Conclusion}
Nearby Seyfert galaxy UGC\,3223 is identified as a new repeat  CL-AGN by our multi-spectroscopy carried out from 
2001 to 2019. The object shows a type transition of $1.5\rightarrow2\rightarrow1.8$ within a period of about 30 years,
and a thorough ``turn-off'' state for at least 10 years. We argue that the type transition of UGC\,3223 not only
could be explained by the disk-wind BLR model, but also suggests a once-dormant and reactive central engine.   
We finally state that our proposed space-based ultraviolet patrol mission (Wang et al. 2019a) is potentially an
effective way for selecting CL-AGN candidates and studying the nature of these objects. 

\acknowledgments

The authors thank all the anonymous referees for their careful review and helpful suggestions that improved the manuscript.
The study is supported by the National Natural Science Foundation of China under grant 11773036, and 
by the Strategic Pioneer Program on Space Science, Chinese Academy of Sciences, grant Nos. XDA15052600 \& XDA15016500.
JW is supported by Natural Science Foundation of Guangxi (2018GXNSFGA281007), and by Bagui Young Scholars Program.
Special thanks go to the staff at Xinglong Observatory as a part of National Astronomical Observatories, 
China Academy of Sciences for their instrumental and observational help. This study is supported
by the Open Project Program of the Key Laboratory of Optical Astronomy, NAOC, CAS. 
This study uses the NASA/IPAC Extragalactic Database (NED), which is operated by the Jet Propulsion 
Laboratory, California Institute of Technology, and the data collected by Wide-field Infrared Survey
Explorer (WISE), which is a joint project of the University of California, Los Angeles, and the Jet Propulsion
Laboratory/California Institute of Technology, funded by the National Aeronautics and Space Administration.

\facilities{NAOC 2.16m telescope}

\software{IRAF (Tody 1986, 1993)}


\begin{thebibliography}{}

\bibitem[Antonucci (1993)]{ant93} Antonucci, R. R. J. 1993, \araa, 31, 473
\bibitem[Bruzual \& Charlot (2003)]{brc03} Bruzual, G., \& Charlot, S. 2003, \mnras, 344, 1000
\bibitem[Cardelli et al. (1989)]{car89} Cardelli, J. A., Clayton, G. C., \& Mathis, J. S. 1989, \apj, 345, 245
\bibitem[Cohen et al. (1986)]{coh86} Cohen, R. D., Rudy, R. J., Puetter, R. C., Ake, T. B., \& Foltz, C. B. 1986, \apj, 311, 135 
\bibitem[Dexter \& Begelamn (2019)]{deb19} Dexter, J., \&  Begelman, M. C. 2019, \mnras, 483, L17  
\bibitem[Denney et al. (2014)] {den14} Denney, K. D., De Rosa, G., Croxall, K., et al. 2014, \apj, 796, 134
\bibitem[Du et al. (2015)]{duh15} Du, P., Hu, C., Lu, K. X., et al. 2015, \apj, 806, 22
\bibitem[Du et al. (2014)]{duw14} Du, P., Wang, J. M., Hu, C., Valls-Gabaud, D., Baldwin, J, A., Ge, J. Q., \& Xue, S. J. 2014, \mnras, 438, 2828
\bibitem[Elitzur \& Ho (2009)]{elh09} Elitzur, M., \& Ho, L. C. 2009, \apjl, 701, 91
\bibitem[Elitzur \& Shlosman (2006)]{els06} Elitzur, M., \& Shlosman, I. 2006,  \apjl, 648, 101  
\bibitem[Fan et al. (2016)]{fan16} Fan, Z., Wang, H. J., Jiang, X. J., et al. 2016, \pasp, 128, 115005
\bibitem[Frederick et al. (2019)]{fre19} Frederick, S., Gezari, S., Graham, M. J., et al. 2019, \apj, 883, 31 
\bibitem[Gezari et al. (2017)]{gez17} Gezari, S., Hung, T., Cenko, S. B., et al. 2017, \apj, 835, 144
\bibitem[Graham et al. (2019)]{gra19} Graham, M. J., Ross, N. P., Stern, D., et al. 2020, \mnras, 491, 4925
\bibitem[Greene \& Ho (2005)]{grh05} Greene, J. E., \& Ho, L. C. 2005, \apj, 630 ,122
\bibitem[Greene \& Ho (2007)]{grh07} Greene, J. E., \& Ho, L. C. 2007, \apj, 670, 92
\bibitem[Husemann et al. (2016)]{hus16} Husemann, B., Urrutia, T., Tremblay, G. R., et al. 2016, \aap, 593, L9
\bibitem[Hutsemekers et al. (2019)]{×} Hutsemekers D., Agis Gonzalez B., Marin F., Sluse D., Ramos Almeida C., Acosta Pulido J.-A., 2019, \aap, 625, A54
\bibitem[Jiang et al. (2016)]{jia16} Jiang, Y. F., Davis, S. W., \& Stone, J. M. 2016, \apj, 827, 10
\bibitem[Kaspi et al. (2005)]{kas05} Kaspi, S., Maoz, D., Netzer, H., Peterson, B. M., Vestergaard, M., \& Jannuzi, B. T. 2005, \apj, 629, 61  
\bibitem[Kaspi et al. (2000)]{kas00} Kaspi, S., Smith, P.S., Netzer, H., Maoz, D., Jannuzi, B.T., \& Giveon, U. 2000, \apj, 533, 631
\bibitem[Kriss (1994)]{krs94} Kriss, G. 1994, Adass, 3, 437
\bibitem[LaMassa et al. (2015)]{lam15} LaMassa, S. M., Cales, S., Moran, E. C., et al. 2015, \apj, 800, 144
\bibitem[Lawrence (2018)]{law18} Lawrence, A. 2018, \nat\ Astronomy, 2, 102
\bibitem[MacLeod et al. (2019)]{mac19} MacLeod, C. L., Green, P. J., Anderson, S. F., et al. 2019, \apj, 874, 8
\bibitem[MacLeod et al. (2010)]{mac10} MacLeod, C. L., Ivezic, Z., Kochanek, C. S., et al. 2010, \apj, 721, 1014 
\bibitem[MacLeod et al. (2016)]{mac16} MacLeod, C. L., Ross, N. P., Lawrence, A., et al. 2016, \mnras, 457, 389
\bibitem[Mainzer et al. (2014)]{mai14} Mainzer, A., Bauer, J., Cutri, R. M., et al. 2014, \apj, 792, 30
\bibitem[Marin et al. (2019)]{mar19} Marin, F., Hutsemekers, D., \& Agis Gonzalez, B. 2019, proceedings of the 2019's annual conference of the SF2A , arXiv/astro-ph:1909.02801
\bibitem[Marziani \& Sulentic (2012)]{mas12} Marziani, P., \& Sulentic, J. W. 2012, \nar, 56, 49
\bibitem[Mathur et al. (2018)]{mat18} Mathur, S., Denney, K. D., Gupta, A., et al. 2018, \apj, 886, 123
\bibitem[Massey et al. (1988)]{mas88} Massey, P., Strobel, K., Barnes, J. V., et al. 1988, \apj, 328, 315
\bibitem[McElroy et al. (2016)]{mc16} McElroy, R. E., Husemann, B., Croom, S. M., et al. 2016, \aap, 593, L8
\bibitem[Nicastro (2000)]{nic00} Nicastro, F. 2000, \apjl, 530, 65
\bibitem[Peterson (2014)]{pet14} Peterson, B. M. 2014, SSRv, 183, 253
\bibitem[Raimundo et al. (2019)]{rai19} Raimundo, S. I., Vestergaard, M., Koay, J. Y., Lawther, D., Casasola, V., \& Peterson, B. M. 2019, \mnras, 486, 123
\bibitem[Ross et al. (2018)]{ros18} Ross, N. P., Ford, K. E. S., Graham, M., et al. 2018, \mnras, 480, 4468 
\bibitem[Ruan et al. (2016)]{rua16} Ruan, J. J., Anderson, S. F., Cales, S. L., et al. 2016, \apj, 826, 188
\bibitem[Rumbaugh et al. (2018)]{rum18} Rumbaugh, N., Shen, Y., Morganson, E., et al. 2018, \apj, 854, 160
\bibitem[Runnoe et al. (2016)]{tun16} Runnoe, J. C., Cales, S., Ruan, J. J., et al. 2016, \mnras, 455, 1691
\bibitem[Schlafly \& Finkbeiner (2011)]{scf11}	Schlafly, E. F., \& Finkbeiner, D. P. 2011, \apj, 737, 103
\bibitem[Shapovalova et al. (2010)]{sha10} Shapovalova, A. I., Popovic, L. C., Burenkov, A. N., et al. 2010, \aap, 509, 106
\bibitem[Shappee et al. (2014]{sha14} Shappee, B. J., Prieto, J. L., Grupe, D., et al. 2014, \apj, 788, 48
\bibitem[Sheng et al. (2017)]{she17} Sheng, Z., Wang, T., Jiang, N., et al. 2017, \apjl, 846, 7
\bibitem[Stern et al. (2018)]{ste18} Stern, D., McKernan, B., Graham, M. J., et al. 2018, \apj, 864, 27
\bibitem[Stirpe (1990)] {sti90} Stirpe, G. M. 1990, \aaps, 85, 1049 
\bibitem[Tody (1986)]{tod86} Tody, D. 1986, SPIE, 627, 733
\bibitem[Tody (1992)]{tod92} Tody, D. 1992, ASPC, 52, 173
\bibitem[Trakhtenbrot et al. (2019)]{tra19} Trakhtenbrot, B., Arcavi, I., MacLeod, C. L., et al. 2019, \apj, 883, 94
\bibitem[Veron-Cetty \& Veron (2006)] {vev06} Veron-Cetty, M. -P., \& Veron, P. 2006, \aap, 455, 773
\bibitem[Wang et al. (2019a)]{wan19a} Wang, J., Liang, E. W., \& Wei, J. Y. 2019a, \pasp, 131, 095001
\bibitem[Wang et al. (2018)]{wan18} Wang, J., Xu, D. W., \& Wei, J. Y. 2018, \apj, 858, 49
\bibitem[Wang et al. (2019b)]{wan19b} Wang, J., Xu, D. W., Wang, Y., Zhang, J. B., Zheng, J., \& Wei, J. Y. 2019b, \apj, 887, 15
\bibitem[Wright et al. (2010)]{wri10} Wright, E. L., Eisenhardt, P. R. M., Mainzer, A. K., et al. 2010, \aj, 140, 1868
\bibitem[Yan et al. (2019)]{yan19} Yan, L., Wang, T. G., Jiang, N., et al. 2019, \apj, 874, 44
\bibitem[Yang et al. (2018)]{yan18} Yang, Q., Wu, X. B., Fan, X. H., et al. 2018, \apj, 862, 109











 













\end{thebibliography}
\end{document}